\documentclass[a4paper,10pt]{article}
\usepackage[utf8]{inputenc}
\usepackage{graphicx,color}
\title{Compact Stars in Vector-Tensor-Horndeski  Theory of Gravity}
\author{ Davood Momeni $^{a}$, Mir Faizal$^{b,c}$, Kairat Myrzakulov $^{a}$,  Ratbay Myrzakulov $^{a}$\\
\\$^{a}$Eurasian International Center for Theoretical Physics\\ and Department of General   \&
Theoretical Physics, \\Eurasian National University, Astana 010008, Kazakhstan.
\\$^{b}$Irving K. Barber School of Arts and Sciences, 
\\ University of British
Columbia - Okanagan,  
3333 University Way,\\  Kelowna,   British Columbia, V1V 1V7, Canada.
\\$^{c}$ Department of Physics and Astronomy, University of Lethbridge,\\
Lethbridge, Alberta, T1K 3M4, Canada.
}
\date{}
\begin{document}

\maketitle

\begin{abstract}
In this paper, we will analyse  a theory of  modified gravity, 
in which the field content of general relativity will be increased to include a vector field. 
We will use  the Horndeski formalism   to non-minimally  couple this  vector field to  the metric.
As we will be using the Horndeski formalism, this theory will not contain Ostrogradsky ghost degree of freedom. 
We will analyse    compact stars using this Vector-Tensor-Horndeski theory. 
\end{abstract}
\section{Introduction}
Even though general relativity is a very  well-tested theory, there is a strong motivation to modify general relativity at large scale.  This is because    
to explain the  accelerating cosmic expansion  in   general relativity, 
a cosmological constant has to be included  \cite{c1, c2, c2a, c4, c5, c6} . Even though the existence of 
such a cosmological constant is predicted from quantum field theories, quantum field theories 
 predict a cosmological constant which is $10^{120}$ times larger than the observed value of the cosmological constant. 
 This has motivated the study of modified theories of gravity, and  
the  Scalar-Tensor theories are such a  modification of general relativity which could  explain accelerating cosmic expansion \cite{s1, t1abcd}. 
 However, these theories in general contain higher derivative terms in the action, and such higher derivative terms   gives rise to 
 Ostrogradsky ghost degree of freedom. These in turn cause instabilities in the theory called the Ostrogradsky instabilities. 
 It is possible to avoid 
 Ostrogradsky ghosts  by using   a theory with galileon symmetry \cite{s2, t2abcd,Jamil:2013yc,Momeni:2014uwa}.    Even though this galileon theory also contains higher 
 derivative terms in the action, the  galilean symmetry ensures that the field equations are only  second-order differential equations.  
 Thus, the galilean symmetry ensures that  
the Ostrogradski ghost instabilities are avoided in these theories. In the decoupling limit, this theory
contains a higher-order derivative interaction known as the cubic galileon. It is possible to construct a theory with quartic and quintic Galileon 
\cite{s12}.   It is necessary  to add couplings of the scalar to curvature tensors,  away from the decoupling limit. This way the Horndeski theory is obtained, and  field equations for both the scalar field  and the metric are again second order differential equations  \cite{h1,h2}. So, Horndeski theory is also free from Ostrogradsky ghosts, and it does not contain the Ostrogradsky instabilities associated with these ghosts. 

This Horndeski formalism is a general formalism and it can be used to analyse a non-minimal coupling of other  fields to the metric. Thus, it is possible to   analyse a Horndeski  coupling of the metric to a vector field \cite{h4,h5}. In this formalism, the vector field is again coupled to the metric using a non-minimal coupling. Furthermore, the field equations for this vector field are again second-order field equations. Thus, the Ostrogradsky ghost terms are avoided even for the Horndeski coupling of a vector field to the metric. As we will be interested in applying such a Vector-Tensor-Horndeski theory to astrophysics, we will assume such a vector field is a fundamental field in nature, which is obtained by increasing the field content of general relativity, and it is not the usual electromagnetic field. This is important as astrophysical objects are neutral and do not have an electric charge. So, the electromagnetic field cannot have a direct effect on the physics of such astrophysical objects. Furthermore, it is known that the gravity couples to the electromagnetic field in the usual way, and there is no reason for it to couple 
to the physical electromagnetic field in a non-minimal way in any astrophysical object. However, if we assume that there exists an astrophysical vector field, which has negligible effect on small scale, then it is possible that such an astrophysical vector field can couple to the metric in a non-minimal way. This coupling of such an astrophysical vector field can have non-trivial effects on the astrophysical phenomena. So,  in this paper, we will analyse the effect of such a non-minimal coupling of an astrophysical   vector field on the physics of a compact star.

It may be noted that such fundamental vector fields have been proposed as a solution to various different physical problems, and  have also been used to     explain many  interesting astrophysical phenomena. Just like the scalar fields, the vector field have also   been used to explain cosmic expansion \cite{21, 21a}, and even the    naturalness problem \cite{22, 22a}. It  may be noted that there the Vector-Tensor theories produce many   non-trivial phenomenological effects which can not be   produced  in the
 Scalar-Tensor theories \cite{12, 12a, 12b, 12c}.   It has been  observed that  anomalies exist in the 
  alignment of the low multipoles of the CMB \cite{16, 16a} and  the hemispherical asymmetry \cite{17}. Such anomalies suggest that there might be a preferential direction in the universe, and this can be explained using such a vector field. Furthermore, such fundamental vector fields have also been used to study inflation \cite{18,19}. The occurrence of higher derivative terms   can produce Ostrogradski ghost instability in such Vector-Tensor theories. However, such Ostrogradski ghost instability can be avoided by using the Horndeski  formalism. 
  Thus, it is interesting to study a Vector-Tensor-Horndeski theory. 
 It may be noted that there are other interesting motivation to introduce such vector fields. This is because there exists a discrepancy between the  predicted and observed dynamics of galaxies ~\cite{rubin1,rubin2,Zwicky}, and   it has  been proposed that the discrepancy can be resolved by increasing the field content of general relativity  \cite{Moffat1, MoffatRahvar1,MoffatRahvar2,BrownsteinMoffat1,MoffatToth, BrownsteinMoffat2}. The field content of this modified theory of gravity  contains such a vector field.  It may be noted that  this theory agrees with  the predictions made by the   modified newtonian dynamics \cite{mond, mond1}.

 As the field content of general relativity contains a vector field, it  has been observed  that  astrophysical black holes in this modification of gravity can mathematically  resemble a Reissner–Nordstrom solution  ~\cite{Moffat5,Moffat6}. However, the   charge of this vector field is generated from mass and not an electromagnetic source \cite{Moffat5,Moffat6}.   As the   astrophysical objects are not charged, and 
 so any coupling of metric to a electromagnetic field cannot be applied to study such astrophysical objects. 
 However, as this  additional vector field is produced from mass and not electric charge \cite{Moffat5,Moffat6},  it is possible that such a field can have direct effect on  astrophysical phenomena. Furthermore, in this theory, there is a certain amount of freedom to choose the action for  vector  field, and it has been demonstrated   that coupling the metric to a 
 non-linear vector field can turn a black hole into a gray hole \cite{mirf}.
 As it has been demonstrated that different  form of the vector field action can produce different  physical results, 
  it is interesting to investigate other forms of coupling of vector field to metric.  Furthermore, as this vector field occur in the field content of general relativity, it will  also be interesting to investigate the astrophysical    application of such a theory. So, in this paper, we will use the Horndeski formalism to couple a  vector field to the metric, and we use this Vector-Tensor-Horndeski theory to analyse a compact star. 
 
%%%%%%%%%%%%%%%%%%%%%%%%%%%%%
\section{ Vector-Tensor-Horndeski Theory}

In this section, we will analyse the main features of a Vector-Tensor-Horndeski theory. 
It has been proposed that by increasing the field content of general relativity certain astrophysical  phenomena  can be explained  
\cite{Moffat1, MoffatRahvar1,MoffatRahvar2,BrownsteinMoffat1,MoffatToth, BrownsteinMoffat2}. Furthermore, for compact stars  this would deform   astrophysical solutions, 
and these deformed solutions would resemble a charged black hole solution \cite{Moffat5,Moffat6}. It is physically important to point out that
the vector field introduced here is not the usual electromagnetic field, but an astrophysical vector field and it has negligible effect at small scale. However, it 
  is expected to change the astrophysical dynamics at larger scale. In this section, we will use the Horndeski formalism to non-minimally couple 
such a vector field to the metric.  Thus, we first introduce as astrophysical vector field $A_\mu$  with the field tensor  
$F_{\mu\nu}=\nabla_{\nu}A_{\mu}-\nabla_{\mu}A_{\nu}$.  We denote the source for such a vector field as 
  $J^{\nu}$ . Now the energy-momentum tensor for this astrophysical vector field can be written as  
  $T_{\mu\nu}=\frac{1}{4\pi}\Big(F_{\mu}^{\alpha}F_{\nu\alpha}-\frac{1}{4}g_{\mu\nu}F^{\alpha\beta}F_{\alpha\beta}\Big)$,
  and we will also denote the energy-momentum tensor for other fields in the theory by $T_{\mu\nu}^M$ .  There is a conserved charge associated with this vector field, 
  as we can write  a divergence free current $\nabla_{\nu}J^{\nu}=0$.  Now  using Horndeski formalism \cite{h4, h5}, we couple this astrophysical vector
  field to the metric as
\begin{eqnarray}
 &&G_{\mu\nu}= 8\pi G(T_{\mu\nu}+\kappa U_{\mu\nu}+T_{\mu\nu}^M), \nonumber \\ &&
\nabla_{\mu}F^{\mu\nu}+\frac{\kappa}{2}\nabla_{\alpha}F_{\beta\gamma}\mathcal{R}^{\nu\alpha\beta\gamma}=4\pi J^{\nu}.
\end{eqnarray}
where $U^{\mu\nu}$ is given by 
\begin{eqnarray}
&&U^{\mu\nu}=\frac{1}{8\pi}\Big(F_{\alpha\beta}F^{\beta}_{\gamma}\mathcal{R}
^{\mu\alpha\nu\gamma}
+\nabla^{\beta}\mathcal{F}^{\mu\alpha}\nabla_{\alpha}\mathcal{F}^{\nu}_{\beta}
\Big)\label{U}
\end{eqnarray}
It may be noted that if $\kappa = 0$, this theory reduced to the usual Vector-Tensor theory. However, we would like to analyse this modified Vector-Tensor-Horndeski theory. The dual tensors can be defined as   
$\mathcal{R}^{\alpha\beta}_{\mu\nu}= \eta^{\alpha\beta\gamma\delta}\eta_{\mu\nu\epsilon\zeta}R^{\epsilon\zeta}_{\gamma\delta}/4
$ and  $\mathcal{F}^{\alpha\beta}= \eta^{\alpha\beta\gamma\delta}F_{\gamma\delta}/2$, where 
  $\eta^{\alpha\beta}_{\gamma\delta}= \delta^{[\alpha\beta]}_{[\gamma\delta]}/ 4!$ is total asymmetric Levi-Civita tensor.   Thus, we can write the Lagrangian for this Vector-Tensor-Horndeski theory as 
\begin{eqnarray}
&&\mathcal{L}=-\frac{R}{16\pi G}+F^{\alpha\beta}F_{\alpha\beta}+\frac{\kappa}{2}F_{\alpha\beta}F^{\gamma\delta}\mathcal{R}^{\alpha\beta}_{\gamma\delta}
\label{L1}.
\end{eqnarray}
This Lagrangian has a non-trivial coupling between  the  astrophysical vector  field and the metric, which was not present in other Vector-Tensor theories of gravity \cite{Moffat1, MoffatRahvar1,MoffatRahvar2,BrownsteinMoffat1,MoffatToth, BrownsteinMoffat2} .

The static spherically symmetric  solutions  for a vector field has been studied using the Horndeski formalism \cite{h4,h5}. 
We will apply this solution to analyse a compact star in this theory, because even though physically, this vector  field is an astrophysical vector field, mathematically, this solution will resemble the  static spherically symmetric  solutions  for a Horndeski vector field   \cite{h4,h5}.  
However, unlike the   electromagnetic fields which cannot have a direct  effect on compact stars, this vector field can change the 
behavior of compact stars. 
Now for this Horndeski   astrophysical vector field    $A_{\mu}$ , we have     $J_{\mu}dx^{\mu}=j(r)dr$,  and for isotropic matter fields,  we can take  a perfect fluid with  $T^{\mu (M)}_{\nu}=\big(\rho,-p-p,-p\big)$. The  static spherically symmetric  metric    for a compact star can be written in the Schwarzschild-Droste coordinates $x^{\mu}=(t,r,\varphi,\theta)$.  Furthermore,  for this astrophysical  vector field, we write  $F_{\mu\nu}dx^{\mu}\otimes dx^{\nu}=f(r) g_{tt}g_{rr}(dt\otimes dr-dr\otimes dt)$.
Here we  choose the units,  such that $c=1$. Now the  metric for this  solution  can be written as 
\begin{eqnarray}
 ds^2=e^{2\psi(r)}dt\otimes dt-e^{2\phi(r)}dr\otimes  dr-r^2d\theta\otimes d\theta-r^2\sin^2\theta d\varphi\otimes d\varphi\label{metric}. && 
\end{eqnarray}
The equation of motion for this solution can be written as  
\begin{eqnarray}
\label{1}&&e^{-2\phi}\Big(\frac{2\phi'}{r}+\frac{e^{2\phi}-1}{r^2}\Big)
=f^2+\frac{\kappa f^2}{r^2}(e^{-2\phi}-1)+8\pi G \rho,
\\&&e^{-2\phi}\Big(\frac{2\psi'}{r}+\frac{1-e^{2\phi}}{r^2}\Big)
=-f^2-\frac{\kappa f^2}{r^2}(3e^{-2\phi}-1)+8\pi G p
\label{2}\\&&e^{-2\phi}\Big(\psi''-\phi'\psi'+\psi'^2+\frac{\psi'-\phi'}{r}\Big)=f^2+8\pi G p\label{3}\nonumber \\ &&  +\frac{\kappa f e^{-2\phi}}{r}    \big(f(\phi'-\psi')-2f'\big)\\&&
f'+\frac{2f}{r}-\frac{\kappa}{r^2}\big(f'(1-e^{-2\phi})+2f\phi'e^{-2\phi}\big)=4\pi j(r)\label{4},
\end{eqnarray}

We can write the hydrostatic equation for matter sector as$\nabla_{\mu}T^{\mu(M)}_{\nu}=0$ for $\nu=r$ , and thus we obtain 
$
 p'+\psi'(p+\rho)=0\label{5}.
$
Now we can write Eqs.  (\ref{1}-\ref{5}) in terms of thermodynamic parameters, and this can be done by 
  redefine the metric function $\phi$ in terms of a mass function $M(r)$ as 
\begin{eqnarray}
e^{-2\phi}=1-\frac{2GM}{r}\label{dM}.
\end{eqnarray}
So, we can obtain  the differential change in mass $dM$, which is the mass   stored in a layer with thickness $dr$  as 
\begin{equation}\label{dM2}
\frac{G dM}{dr}=\frac{1}{2}\Big[1-e^{-2\phi}(1-2r\phi')\Big],
\end{equation}
where $M$ is the mass of the compact object. Now   make the equations dimensionless by  expressed them in terms of $\{\frac{dp}{dr},\frac{dM}{dr},\rho,p\}$, and  by using  $M\to m M_{\odot}$, $r\to r_{g} r$, $\rho\to \rho M_{\odot}/r_{g}^3$, $p\to p M_{\odot}/r_{g}^3$ and $R\to R/r_g^2$. Here $r_{g}=G_NM_{\odot}=1.47473 \mbox{km}$ , $M_\odot$ is   the mass of the Sun, and  
\begin{eqnarray}
\frac{d\phi}{dr}=\frac{m}{r^2}\frac{1-\frac{r}{m}\frac{dm}{dr}}{\frac{2m}{r}-1}\label{dm}.
\end{eqnarray}
Thus, we obtain 

\begin{eqnarray}
\label{11}&&\frac{2m}{r^3}(1-\frac{r}{m}\frac{dm}{dr})-\frac{2m}{r^3}
+r_g^2f^2-\frac{2m\kappa f^2}{r^3}=8\pi \rho,
\\&&
\label{22}\frac{2p'}{r(p+\rho)}(1-\frac{2m}{r})+\frac{2m}{r^3}-r_g^2f^2-\frac{2\kappa f^2}{ r^2}(1-\frac{3m}{r})=8\pi p
\\\label{33}&&\Big(\frac{p'}{p+\rho}\Big)'-\frac{m}{r^2}\frac{1-\frac{r}{m}\frac{dm}{dr}}{\frac{2m}{r}-1}\frac{p'}{p+\rho}-\Big(\frac{p'}{p+\rho}\Big)^2+\frac{p'}{r(p+\rho)}\nonumber \\ && +\frac{m}{r^3}\frac{1-\frac{r}{m}\frac{dm}{dr}}{\frac{2m}{r}-1}+\frac{f^2 r_g^2}{1-\frac{2m}{r}}\\&&\nonumber+\frac{\kappa f }{r}\big(f(\frac{m}{r^2}\frac{1-\frac{r}{m}\frac{dm}{dr}}{\frac{2m}{r}-1}+\frac{p'}{p+\rho})-2f'\big)=\frac{8\pi p}{1-\frac{2m}{r}}\\\label{44}&&
f'+\frac{2f}{r}-\frac{\kappa r_g ^{-2}}{r^2}\big(\frac{2mf'}{r}-\frac{2mf}{r^2}(1-\frac{r}{m}\frac{dm}{dr})\big)=4\pi r_gj(r),
\end{eqnarray}

We will analyse a compact start by solving Eqs.  (\ref{11}-\ref{44}). Furthermore, we will also   use the  equation of state   $p=p(\rho)$ and a specific  form of $j(r)$ to obtain such   solutions. 
The function $m(r)$  is important in  analyzing the geometric mass  inside a sphere of radius $r$. 
This is because we can use Eq.  (\ref{dM2}) , and write 
\begin{eqnarray}
&&m'(r)=\frac{1}{2}\Big[r(1-e^{-2\phi})\Big]'. 
\end{eqnarray}
Furthermore, from Eq.  (\ref{1}), we obtain 
\begin{eqnarray}
&&m'(r)=4\pi r^2\rho-r^2f^2\Big[1+\frac{\kappa}{r^2}(e^{-2\phi}-1)\Big]\label{MR3}.
\end{eqnarray}
It may be noted in  the absence of the Horndeski field,  $f=0$, the Eq. (\ref{MR3}) reduces to the usual  form $m'(r)=4\pi r^2\rho$ in general relativity. Now by integrating Eq. (\ref{MR3}), we obtain 
\begin{eqnarray}
&&m(R)\equiv M=\int_{0}^{R} \Big(4\pi r^2\rho-r^2f^2\Big[1+\frac{\kappa}{r^2}(e^{-2\phi}-1)\Big]\Big) dr\label{M-R}.
\end{eqnarray}
Now using this result, we can obtain 
\begin{eqnarray}
\label{TOV}&&
\frac{2p'}{r(p+\rho)}(1-\frac{2m}{r})+\frac{2m}{r^3}+\Big[\frac{8\pi \rho+\frac{2m}{r^3}(1-\frac{r}{m}\frac{dm}{dr})-\frac{2m}{r^3}}{r_g^2-\frac{2m\kappa }{r^3}}\Big]  \nonumber \\ && 
\Big(r_g^2+\frac{2\kappa}{ r^2}(1-\frac{3m}{r})\Big)=8\pi p
\end{eqnarray}
Now the   equation of state for this system can be written as   $p=p(\rho)$, It is possible to 
 integrate this equation of state to obtain the behavior of this system in this Vector-Tensor-Horndeski theory. 
The boundary conditions used for analyzing this system in 
this Vector-Tensor-Horndeski theory are similar to those used in the 
  Einstein gravity,    $m(0)=0,\rho(0)=\rho_c,$
and $ p(r_0)=0$. So, the radius of the compact star can be taken  
to be $r_0$ , such that   the pressure vanishes.  
Now we can analyse compact stars using this formalism.  In astrophysics,  the term compact star  is used to 
collectively refer to white dwarfs, 
neutron stars, and black holes.  They are described by Tolman-Oppenheimer-Volkoff  equations \cite{t1, t2}. 
In this section, we will analyse the effect of the Horndeski astrophysical vector field on the physics
of such compact stars.  
%%%%%%%%%%%%%%%%%%%%%%%
\section{The Gravitational Binding Energy}
The  gravitational binding energy  of a system is the minimum energy that must be added to that system for it to stop being a 
 a gravitationally bound system.
It is important to analyze the gravitational binding energy of compact stars, such as the neutron star, and such an analysis has been done 
using  general relativity \cite{g0, g1, g2}. 
So, in this section, 
we shall analyze the effect of the astrophysical Horndeski field on the gravitational binding energy of a compact star. 
Thus, using the  definition of the mass function  (\ref{M-R}), the total mass of the matter distribution of this system can be represented by 
\begin{eqnarray}
&&m_{ADM}=4\pi\int_{o}^{\infty}dr \rho(r)r^2+\Delta m_{ADM}. 
\end{eqnarray}
This mass is   the Arnowitt-Deser-Misner (ADM) mass 
and in the Horndeski-Vector-Tensor theory. 
It may be noted that  $\Delta m_{ADM}$ is the corrections to the usual  ADM mass produced by the  the Horndeski vector field, 
\begin{eqnarray}
&&\Delta m_{ADM}=-\int_{0}^{\infty} \Big(r^2f^2\Big[1+\frac{\kappa}{r^2}(e^{-2\phi}-1)\Big]\Big) dr.
\end{eqnarray}
Now we can define  the density inside a proper volume element $\sqrt{-g}d^3x$ as the proper mass,
\begin{eqnarray}
&&M_{pr}=4\pi \int_{0}^{\infty}dr \rho(r) e^{\phi(r)}r^2\label{proper}
\end{eqnarray}
We interpret the difference between the proper (\ref{proper}) and the total mass (\ref{M-R})   as the gravitational binding energy
\begin{eqnarray}
&&E_b=M_{pr}-m>0.
\end{eqnarray}
In general relativity, due to the absence of vector or Horndeski fields, $f=0$ and $\Delta m_{ADM}=0$. 
However, as we have a   non-minimal coupling to the Horndeski field, the ADM mass is corrected by a finite $\Delta m_{ADM}\neq 0$. 

To analyze the effect of the astrophysical vector field on compact stars, we need to solve the 
 modified Tolman-Oppenheimer-Volkoff  equation, Eq. (\ref{TOV}). However, to solve this equation, we need 
 to use the  equation of state for  the interior structure of the star. 
 It is possible to write this equation of state using  the 
 central density, $\rho(r=0)=\rho_c$,  as a free parameter. 
 Thus, we can  obtain the mass and  radius of  a star by fixing the central density. 
 This will correspond to choosing  a single point in the mass-radius  diagram for the star. 
 As this system is described by a single parameter,  we can obtain the 
 full  mass-radius  curve of the star  by varying 
$\rho_c$. It is possible to study the  
inner structure of the compact stars, such as the    neutron stars \cite{poly12}. 
This is because it possible to use microscopic many-body simulations to numerically analyze  equation of state for such 
compact stars.  The equation of state for compact stars   can be obtained using a  
 mean-field theoretical description of such a system     \cite{eos1,eos2,eos3}. 
 In fact, it is possible to describe the  equation of state for a neutron star using the 
 nucleon-nucleon interaction. The equation of state for such a star is 
 is  a polytrope  equation of state \cite{poly12}, 
\begin{eqnarray}
&&p=k\rho^{\gamma}
\end{eqnarray} 
The value of $k$ can be   taken to be  $k\approx 2.0\times10^5 \frac{cm^5}{ gs2}$ when $\gamma = 2$ \cite{poly}.
We can now use this polytrope equation of state to analyze the effect of Horndeski field on a neutron star.
%%%%%%%%%%%%%%%5

\begin{figure}[!htb]
\includegraphics[width=8cm]{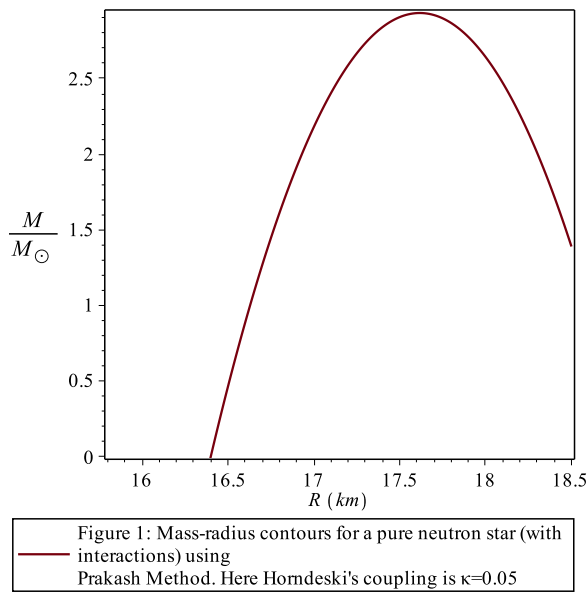}
\caption{Mass-radius contours for a pure neutron star (with interactions) using
Prakash Method. Here Horndeski's coupling is $\kappa=0.05$}
\label{kappa05}
\end{figure}
%%%%%%%%%%%%%%%%%%%%%%%%%%%%%%%%%%
\begin{figure}[!htb]
\includegraphics[width=8cm]{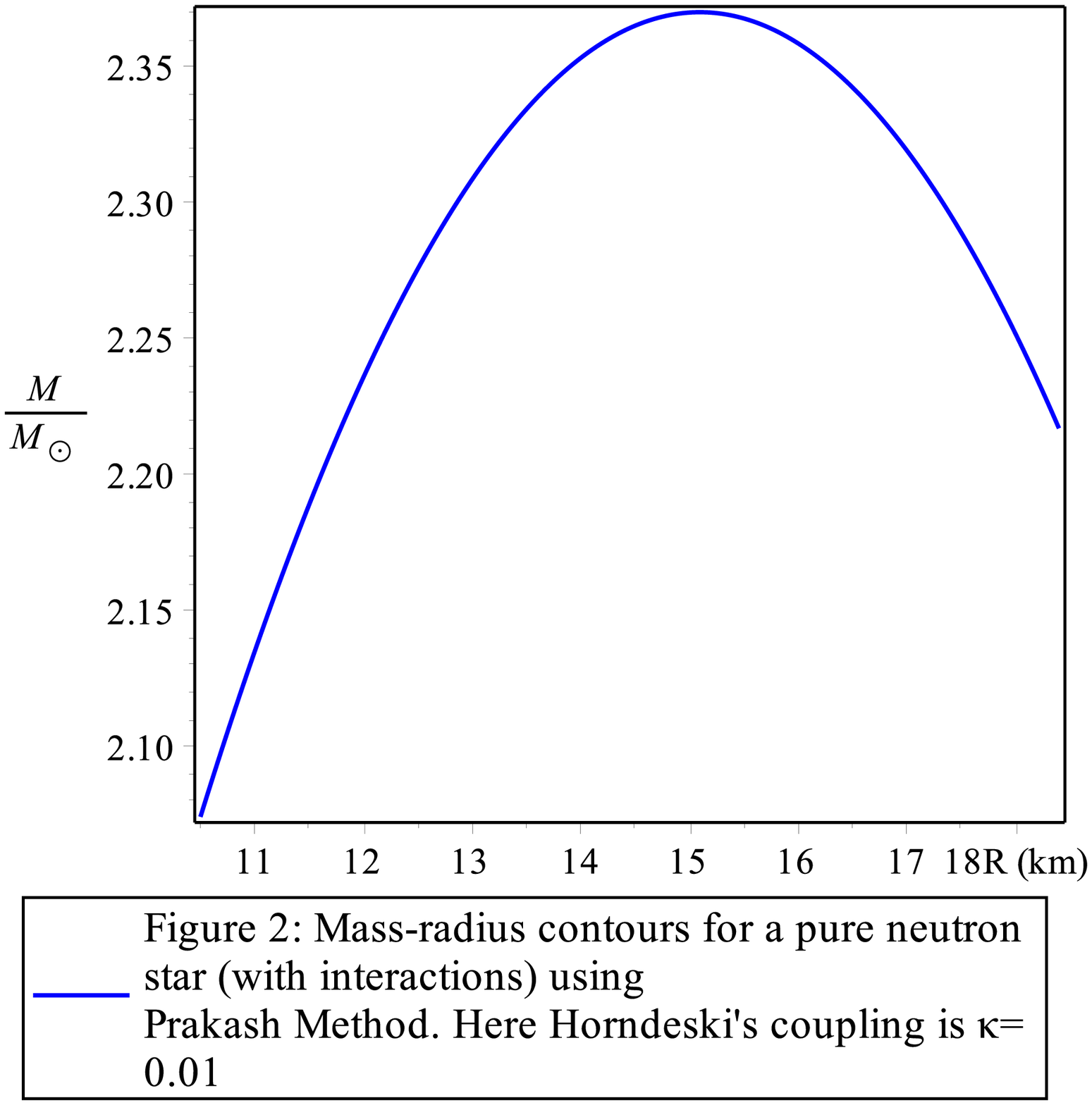}
\caption{Mass-radius contours for a pure neutron star (with interactions) using
Prakash Method. Here Horndeski's coupling is $\kappa=0.01$}
\label{kappa01}
\end{figure}
%%%%%%%%%%%%%%%%%%%%%%%%%%%%%%%%

\begin{figure}[!htb]
\includegraphics[width=8cm]{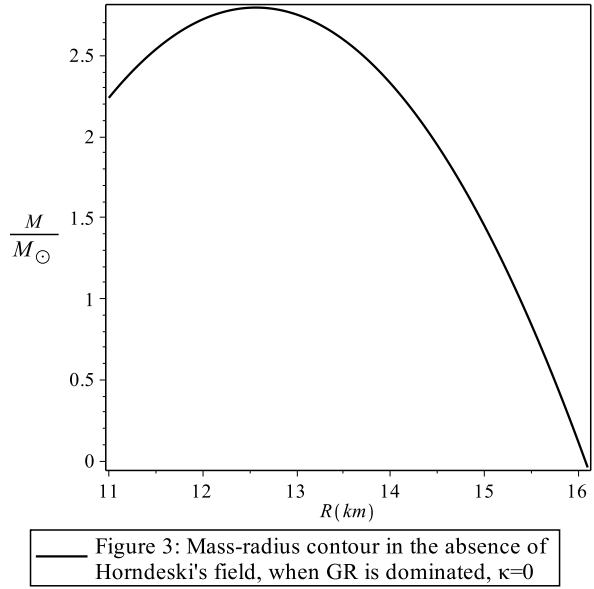}
\caption{Mass-radius contour in the absence of Horndeski's field, when general relativity is dominated, $\kappa=0$.   }
\label{kappa0}
\end{figure}

%%%%%%%%%%%%%%%%%%%%%%%

\begin{figure}[!htb]
\includegraphics[width=8cm]{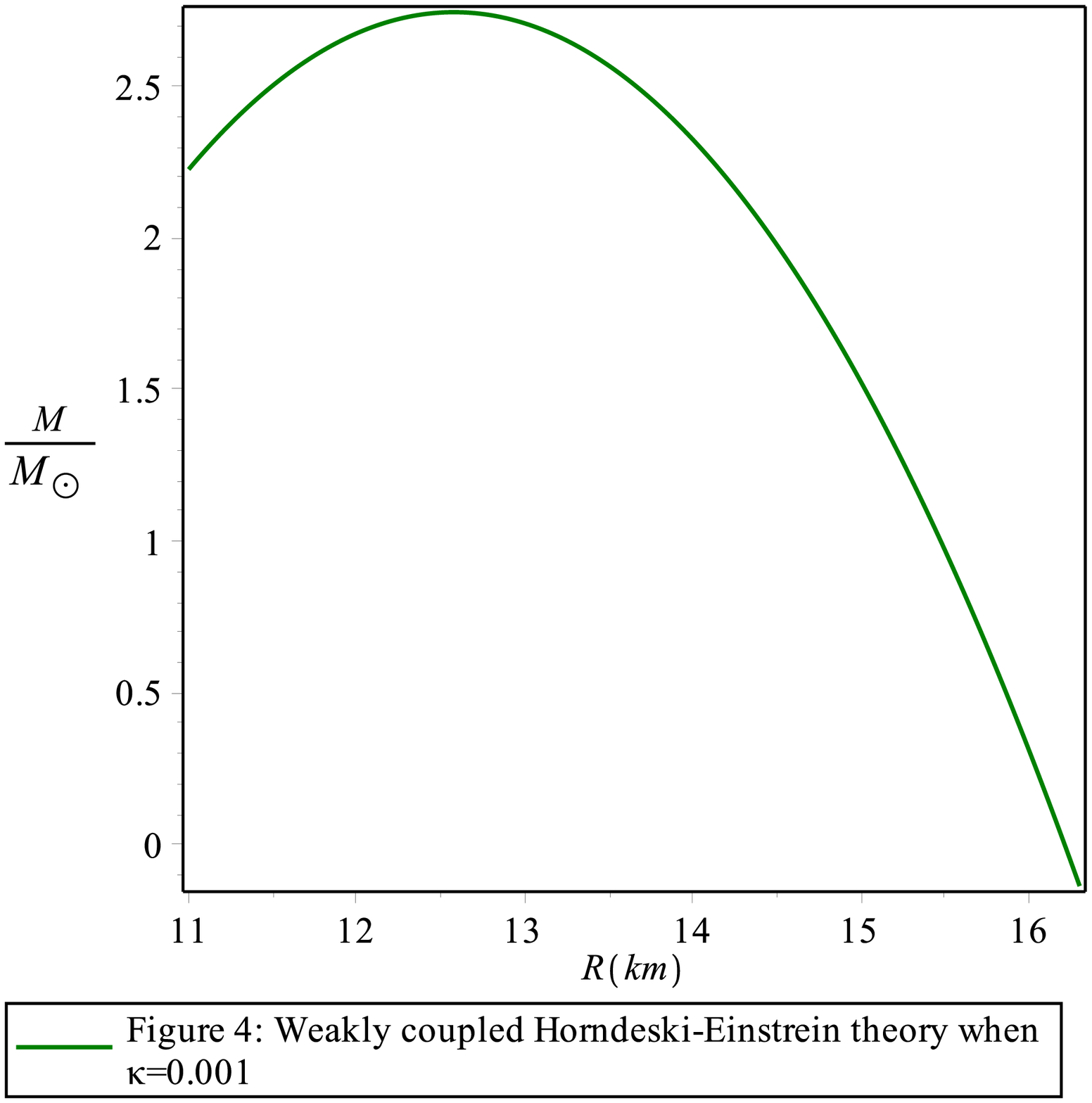}
\caption{Mass-radius contour when Horndeski's field is weakly coupled, $\kappa=0.001$.   }
\label{kappa001}
\end{figure}
%%%%%%%%%%%%%%%%%%%%%55

 We have used an  adaptive step-size
Runge-Kutta method  for analyzing a non-linear-integro-differential equation   \cite{method},\cite{Oliveira:2015lka}. This is because 
the  modified Tolman-Oppenheimer-Volkoff equation 
is a non-linear-integro-differential equation. 
{  In Figure. (\ref{kappa01}), blue line represents the  Vector-Tensor-Horndeski  theory for $\kappa=0.01$. It  
predicts the existence of a typical neutron star with mass around $2M_{\odot}$, and a radius around  $10 -15 Km$.}
Thus, this case is not physical. In fact, this case can also be used to set a bound on the 
strength of the coupling of this astrophysical vector field to general relativity. 
{  In Figure. (\ref{kappa0}), the black line represents the standard general relativity, $\kappa=0$.}
The mass-radius diagram corresponding to this case is given in 
Fig.  (\ref{kappa05})-(\ref{kappa001}).
In Fig. (\ref{kappa05}), the red line represents the Vector-Tensor-Horndeski  theory with 
$\kappa= 0.05$. It does not predict the   
existence of a typical neutron star with mass  around 
$2M_{\odot}$, and with a radius around  $10-15 Km$.  
It may be noted that for $\kappa < 0$,  the numerical evaluation does not converge.
Furthermore, a stable stellar configurations does not exist for this case, as the 
 hydrostatic equilibrium equations  are unstable. 
 The mass-radius profile   for a 
weakly coupled Vector-Tensor-Horndeski  theory ($\kappa= 10^{-3}$) is plotted in Fig.  (\ref{kappa001}).  
 We observed that this profile around the maximum with  mass is similar to the profile obtained in
 the general relativity. 
 However, for  large radius, equilibrium configurations in the Vector-Tensor-Horndeski  theory are more massive 
 than in general relativity.

%%%%%%%%%%%%%%%%%%%%%%%%%%%% 
%%%%%%%%%%%%%%%%%%%%%%%%%%%%%%%
\section{ Density Profile}

It is possible to analyze a compact star to be represented by a constant density. 
It may be noted that the compact stars with constant density have been analysed in general relativity
\cite{co} (see \cite{Schlogel:2016xtf} for a comprehensive review). 
In general relativity the Tolman-Oppenheimer-Volkoff  equation admits an analytical solution,
and this is obtained by  imposing $p(r = r_0) = 0$. Here  
 the central pressure $p_c = p(r = 0)$ predicted by general relativity  becomes infinite 
 for  the critical mass $M_{cr}=4/9 m_{pl}^2 r_0$ \cite{Buchdahl}.  
So, it is possible for stars with $M> M_{cr}$ to   indicate 
a deviation from general relativity \cite{bound}.
Thus, it is interesting to analyze the effect of a Horndeski vector  field for such a system. 
We can first analyse  a uniform mass density called as top-hat density profile inside the star,
\begin{eqnarray}
&&\rho=\cases{\rho_{{0}}&$r<r_{{0}}$\cr 0&$r_{{0}}\leq r$\cr}
\end{eqnarray}
Here we shall now analyse such a system using modified  Vector-Tensor-Horndeski theory.
Now for this model and for simplicity, we also choose  $j(r)=0$.  So, the    unknown functions for this system are    
$\{m(r),f(r)\}$, and the   metric function is  $\psi=\psi_0$. Exact solution for the Vector-Tensor-Horndeski theory, for this system can be written as 
\begin{eqnarray}
&&
m( r ) =\frac{4}{3}\,\pi \,\rho\,{r}^{3}+\,{\frac {2\pi \,\rho
\,{r_{{g}}}^{2}{r}^{5}}{15 \kappa}}+O \left( {r}^{6} \right) \label{m}
\end{eqnarray}
Now using  (\ref{m}) and  Eq. (\ref{dM}), we obtain 
\begin{eqnarray}
&&
\phi \left( r \right) =-\frac{1}{2}\ln  \left( 1+\frac{8}{3}\,\pi \,\rho\,{r}^{2}-{\frac {4}{15}}\,{\frac {\pi \,
\rho\,{r}^{4}{{ r_g}}^{2}}{\kappa}}+O \left( {r}^{5} \right) 
 \right) 
 \label{phi}
\end{eqnarray}
The  metric of this compact star can be written as 

\begin{eqnarray}
 ds^2&=& e^{2\psi_0}dt\otimes dt-r^2d\theta\otimes d\theta-r^2\sin^2\theta d\varphi\otimes d\varphi\nonumber \\ 
 && -\frac{ dr\otimes  dr}{1+\frac{8}{3}\,\pi \,\rho\,{r}^{2} -{\frac {4}{15}}\,{\frac {\pi \,
\rho\,{r}^{4}{{\it rg}}^{2}}{\kappa}}+O \left( {r}^{5}\right)}  \label{metric}.
\end{eqnarray}

Now using this mass profile,  we can  integrate Eq, (\ref{44}), and obtain 
\begin{eqnarray}&&
f \left( r \right) =f_{{0}}+O \left(  \left( r-R
 \right) ^{2} \right) \\&&\nonumber+{\frac { \left( 30\,f_{{0}}{r_{{g}}}^{2}-
80\,\pi \,\rho\,f_{{0}}\kappa-16\,\pi \,\rho\,f_{{0}}{r_{{g}}}^{2}{R}^
{2} \right)  \left( r-R \right) }{-15\,R{r_{{g}}}^{2}+40\,\pi \,\rho\,
R\kappa+4\,\pi \,\rho\,{R}^{3}{r_{{g}}}^{2}}}
\end{eqnarray}
Here we have  assumed that this astrophysical  vector  field satisfies the following initial condition on the surface of star $r=R$,
$
f(R)=f_0. $ As the original function is corrected by a $\kappa$ dependent terms, it can be argued that such  corrections are produced by 
the Horndeski vector field. As such a coupling between gravity and Horndeski vector field is constraint by experimental data, it is possible 
to use the astrophysical data to constraint such a coupling.

%%%%%%%%%%%%%%%%%%%%%%%%%%  
We analyzed a compact star  with a constant  density. However, for real stars, we expect that the density to be a  function of $r$. 
So, it is  interesting to analyse various different functional dependence of $m(r)$, and analyse the effect of Horndeski vector field on 
a system described by such functions. This is important to demonstrate that the dependence of the system on Horndeski  vector field is not a special  feature of 
a system with a constant density. 
 Now we can take a simple  function  $m(r)$ of $r$, such as 
   $m(r)=M_0 \ln(r)$, to demonstrate that the Horndeski vector field effects the systems in which the density is not a constant.  
   Now for such a system, we obtain 
\begin{eqnarray}&&
f(r)=f_{{0}}+O ( ( r-R) ^{2}) +\,{\frac {2f_{{0}} ( {R}^{3}{r_{{g}}}^{2}+\kappa\,M_{{0}}\ln 
( R ) -\kappa\,M_{{0}} ) ( r-R ) }{R
( -{R}^{3}{r_{{g}}}^{2}+2\,\kappa\,M_{{0}}\ln ( R ) 
) }}
\end{eqnarray}
The mass density of this system can be obtained from Eq. (\ref{11}), and it is given by 
\begin{eqnarray}&&
\rho(r)=
\,{\frac {{f_{{0}}}^{2} \Sigma(-{r_{{g}}}^{2}{r}^{3}+2\,m\kappa) }{8{r}^{3}{R}^{2} \left( -{R}^{3}{r_{{g}}}^{2}+2\,
\kappa\,M_{{0}}\ln  \left( R \right)  \right) ^{2}\pi }}
\end{eqnarray}
where we have defined $\Sigma$  as 
\begin{eqnarray}&&
\Sigma^{1/2}= -3\,{R}^{4}{r_{{g}}}^{2}+2\,{R}^{3}{
r_{{g}}}^{2}r+2\,\kappa\,M_{{0}}\ln  \left( R \right) r-2\,\kappa\,M_{
{0}}r+2\,\kappa\,M_{{0}}R.
\end{eqnarray}
The  pressure for this system can also be obtained, and it is given by 
\begin{eqnarray}&&
p(r)={\frac {X\,\rho\, \left( r-R \right) }{R \left( -{r_{{g}}}^{2}
{R}^{3}+2\,M_{{0}}\ln  \left( R \right) \kappa \right)  \left( -R+2\,M
_{{0}}\ln  \left( R \right)  \right) }}\\&&\nonumber+O \left(  \left( r-R \right) ^
{2} \right)
\end{eqnarray}
where we have defined $X$ as 
\begin{eqnarray}&&
X=-M_{{0}}\ln  \left( R \right) {r_{{g}}}^{2}{R}^{3}+2\,{M_{{0
}}}^{2} \left( \ln  \left( R \right)  \right) ^{2}\kappa \\&&\nonumber-4\,\pi \,\rho
\,{R}^{6}{r_{{g}}}^{2}-8\,\pi \,\rho\,{R}^{4}\kappa+24\,\pi \,\rho\,{R
}^{3}M_{{0}}\ln  \left( R \right) \kappa+M_{{0}}{r_{{g}}}^{2}{R}^{3} \\&&\nonumber+2
\,M_{{0}}\kappa\,R-6\,{M_{{0}}}^{2}\ln  \left( R \right) \kappa
\end{eqnarray}
Thus, the pressure of this system is effected by the Horndeski vector field. This is because the pressure 
of this system 
 is  corrected by terms proportional to $\kappa$, and so this system is 
effected by the Horndeski vector field. 

 It may be noted that it is possible to take other functions describing $m(r)$, and analyse 
the pressure of the compact stars using those functions. This procedure can be repeated for those functions. 
 It is expected that the pressure in such systems will also depend on the Horndeski vector field. 
 So, for example,  we can take another form of the function  $m(r)=M_0r\ln(r)+M_1r+M_2r^2$, and demonstrate that the system 
 described by this function will also be effected by the Horndeski vector field. Thus, using this function, we obtain 
 \begin{eqnarray}&&
f \left( r \right) =f_{{0}}   +O \left(  \left( r-R \right) ^{2} \right)\\&&\nonumber
- \left( \,{\frac {2f_{{0}}}{R}}+\frac{2\,\kappa}{{R}^{4}{r_{{g}}}^{2}}
\, \left( M_{{0}}R\ln  \left( R \right) +M_{{2}}{R}^{2}+M_{{1}}R
 \right)A
 \right)\left( r-R \right), 
\end{eqnarray}
 where 
 \begin{eqnarray}A= 
  \frac{f_{{0}}(1-{\frac {R \left( 2\,M_{{2}}R+M_{{1}}+M_{{0}}+M_{{0}}\ln  \left( R
 \right)  \right) }{M_{{0}}R\ln  \left( R \right) +M_{{2}}{R}^{2} +M_{{
1}}R}}
)}{ \left( 1-2\,{\frac {\kappa\, \left( M_{{0
}}R\ln  \left( R \right) +M_{{2}}{R}^{2}+M_{{1}}R \right) }{{R}^{3}{r_
{{g}}}^{2}}} \right)}  
 \end{eqnarray}
 We observe the this system is again corrected by terms proportional to $\kappa$, and so 
 the physics of compact starts is effected by the Horndeski vector field.
 This function can be used to obtain the mass density and pressure of the compact star.
 As the original function were corrected by the Horndeski vector field, the mass density and 
 pressure of the compact star in this theory would again depend on the Horndeski vector field. 
 These results can be compared with experimental data, and 
 bounds on the existence of such an astrophysical vector field can be thus obtained. 
 Thus, we have analyzed compact stars in a Vector-Tensor-Horndeski theory, and observed that the dynamics of this system 
 is corrected by terms which are proportional to the coupling constant of the Horndeski field. 

 %%%%%%%%%%%%%%%
\subsection{Astrophysical Monopole}
 It may be noted that as we are using an astrophysical vector field, it is possible that such a vector field will also contain monopoles.
 Furthermore, as this vector field will have negligible effect at small distance, we cannot rule out the existence of such monopoles in this vector field.
 These monopoles can effect the astrophysical phenomena, and have a direct effect on the physics of compact stars.  Thus, if we assume  a monopole with charge $Q$ is
 located at $r=0$, then  we can write  $j(r)={Q}/{r^2}$. The   exact solution for this system can now be written as 
\begin{eqnarray}&&
f \left( r \right) =f_{{0}}\\&&\nonumber+ \left( {\frac {30\,f_{{0}}{r_{{g}}}^{2}-80\,\pi \,\rho\,f_{{0
}}\kappa-16\,\pi \,\rho\,f_{{0}}{r_{{g}}}^{2}{R}^{2}}{-15\,R{r_{{g}}}^
{2}+40\,\pi \,\rho\,R\kappa+4\,\pi \,\rho\,{R}^{3}{r_{{g}}}^{2}}}+\,{
\frac {4\pi \,r_{{g}}Q}{{R}^{2}}} \right)  \left( r-R \right)\nonumber \\ &&  +O
 \left(  \left( r-R \right) ^{2} \right) 
\end{eqnarray}

As the constant density solutions have been studied in general relativity \cite{co}, it is interesting to analyse different  limits of such solutions. 
 It is possible to take a constant density solution,   $\rho(r)=\rho_0 $ with $j(r)=0$, and the analyse the effect of Horndeski vector field 
 using  $f(r)={f_0}/{r}$, where $f_0$ is a constant. Now using this form of $f(r)$, we can obtain 
 \begin{eqnarray}&&
m \left( r \right) =\,{\frac {{r_{{g}}}^{2}{r}^{3}}{2\kappa}}+c{r}^{
2}
\end{eqnarray}
Now using  this mass profile we can integrate Eq. (\ref{TOV}), and obtain the pressure  as 
\begin{eqnarray}&&
p \left( r \right)=\,{\frac {\rho\,\Delta\, }{2Rc\kappa\, \left( -
\kappa+2\,Rc\kappa+{R}^{2}{r_{{g}}}^{2} \right) }} \left( r-R \right)+O \left(  \left( r-
R \right) ^{2} \right) 
\end{eqnarray}
where we have defined $\Delta$ as 
\begin{eqnarray}
\Delta&=& -12\,{R}^{2}c\kappa\,{r_{{g}}}^{2}+8\,\pi \,\rho\,\kappa\,{R}^{3}{r_{{
g}}}^{2}+24\,\pi \,\rho\,{\kappa}^{2}{R}^{2}c \nonumber \\ && -10\,R{c}^{2}{\kappa}^{2}
-8\,\pi \,\rho\,{\kappa}^{2}R-3\,{R}^{3}{r_{{g}}}^{4}+4\,c{\kappa}^{2}
+3\,R{r_{{g}}}^{2}\kappa. 
\end{eqnarray}
So, the existence of astrophysical monopole  from Horndeski vector field can correct the pressure  of a compact star. 
Thus, astrophysical monopoles can have interesting effects on the physics of compact stars in a Horndeski-Vector-Tensor 
theory of gravity. 
As the monopoles have not been detected in the  electromagnetic vector field, there are strong constrains of including the effects of 
such monopoles in physical systems. However,  the Horndeski vector field is different from electromagnetic vector field, 
so the constraint on the electromagnetic vector field 
from the absence of electromagnetic  monopoles do  not apply to such Horndeski vector fields, and hence it is possible that such 
vector fields can change the physics of compact stars. 

%%%%%%%%%%%%%%%%%%%%%%%
\section{Conclusion}
In this paper, we have  analysed  a theory of  modified gravity. In this theory, the field content of general relativity was increased 
to include an astrophysical vector field. We have used  the Horndeski formalism   to non-minimally  couple this  astrophysical vector field to  the metric. 
As we  have used the Horndeski formalism, this theory did  not contain any Ostrogradsky ghost degree of freedom. 
We will analysed a  compact star using this Vector-Tensor-Horndeski theory.  Thus, we analyzed the effect of such a Horndeski vector field 
on the gravitational binding energy.
We used a polytrope  equation of state for 
analyzing a neutron star in Horndeski-Vector-Tensor 
theory of gravity. 
We analysed this system using an 
adaptive step-size Runge-Kutta method \cite{Oliveira:2015lka}. 
We also  analysed various different cases for  this system,
and obtained the mass density and pressure for the compact stars corresponding to those cases. 
It was demonstrated that the Horndeski changed  the physics of this system for these different cases.
It would be interesting to compare the results of this
paper to experimental data, and thus obtain  bounds on the  Vector-Tensor-Horndeski theory.
Finally, we proposed that it is possible for a monopole to exist in such a modified theory of gravity. 
As no monopole has been detected in the electromagnetic field, there are strong constrains on the existence 
of monopoles in such a theory. However,  the Horndeski vector field was not constraint by the constraint  on the electromagnetic vector field,
and so we analyzed the effects of an astrophysical
monopole on the compact stars. It was demonstrated 
that such an astrophysical vector field would 
change the pressure of the compact star.

It may be  noted that it is possible to analyse 
compact stars with an anisotropy. In fact,   multipole moments for compact stars with anisotropic pressure have been 
studied \cite{2k}.   
The 
 compact stars with    anisotropy have been studied using a  modified Tolman-Oppenheimer-Volkoff equation \cite{1k}. 
 It has been observed that the  pressure anisotropy can effect  the surface tension
 of these stars. This is because the    anisotropy   decreases the value of the surface tension. It would be interesting to introduce such an anisotropy using a 
 vector field. Furthermore, this vector field can be coupled non-minimally to the metric using Horndeski formalism. It would be interesting to analyse the 
 phenomenological effects of such a model.  The vector fields have been used to modify general relativity, and it has been possible to use this modified 
 theory of gravity to obtain the correct dynamics of galaxies without the need for dark matter 
 \cite{Moffat1, MoffatRahvar1,MoffatRahvar2,BrownsteinMoffat1,MoffatToth, BrownsteinMoffat2}.  It would be interesting to 
 analyse the dynamics of galaxies using a Horndeski vector field. It might be possible to obtain the correct dynamics of galaxies by suitable modifying such a theory, 
 and by possible adding other fields to it. It would also be interesting to analyse the effects of such a theory on inflation. This is because vector  fields have
 been used to study inflation \cite{18,19}, and it would be interesting to repeat this calculation using the Horndeski formalism. 

It is possible to make observations on 
a neutron star, and these observations can be compared 
with the predictions of the Vector-Tensor-Horndeski theory. 
This can be used to both verify the existence of a Horndeski 
vector field, if such effects are detected. However, if 
such effects are not detected then it can be used to 
set bounds on the strength of the coupling of such 
a astrophysical vector field to general relativity. 
It may be noted that as the $\kappa= 0.05$ does not predict the existence of a 
 neutron star with a  typical  mass and radius, so this case does not fit the experimental data. Thus, this 
 case is not physical. In fact, this case also establish a bound on the strength of coupling parameter 
 of the Horndeski vector field to general relativity. It may be noted that this coupling does change the behavior 
 of mass-radius diagram. Such a mass-radius diagram of a neutron star can be observed, and the observations can be 
 compared with this analysis. This can be used to test the existence of such a Horndeski astrophysical vector field, 
 and also establish a bound on the strength of 
 coupling of such a field. It may be noted that it is possible to obtain such experimental data for  neutron stars using 
 gravitational lensing \cite{lense1, lense2}.  However, it is also important to analyze the effect of Horndeski vector 
 field on the gravitational lensing for such an analysis.
 It is possible to use the burst oscillations 
 in the X-ray flux to obtain the observational
 behavior of mass-radius of a neutron star \cite{xray}. 
 It is also possible to use quescent neutron stars to 
make such observations on a neutron star \cite{quescent}. 
 It would be interesting to 
 analyze the effect of the Horndeski vector
 field on neutron stars using such data. 

 It would be interesting to analyze other effects of this Horndeski vector field which can be observed using compact stars. 
 The    accretion in the Reissner–Nordstrom spacetime has already been studied \cite{reis}. It was observed that 
 the electromagnetic  field can have interesting effect on such an accretion around a compact star. In fact, the 
 effect of a monopole on the accretion has also been studied \cite{reis1}. As the astrophysical objects are not changed,   such 
 systems can not be physically realized. However, in this paper, we have proposed that a Horndeski astrophysical vector field can couple 
 to general relativity, and the bounds on the strength of such an coupling can be obtained from observations. 
 Such an astrophysical vector field will also change the   accretion around compact stars. It would be interesting 
 to analyze the   accretion in this Vector-Tensor-Horndeski theory. It would also be interesting to compare the result 
 thus obtained to observations, and test the Vector-Tensor-Horndeski theory.


\begin{thebibliography}{99}





\bibitem{c1}A. G. Riess et al., Astron. J. 116, 1009 (1998)
\bibitem{c2}S. Perlmutter et al., Nature 391, 51 (1998)
\bibitem{c2a}A. G. Riess et al., Astron. J. 118, 2668 (1999)
\bibitem{c4} S. Perlmutter et al., Astrophys. J. 517, 565 (1999)
\bibitem{c5} A. G. Riess et al., Astrophys. J. 560, 49 (2001)
\bibitem{c6} J. L. Tonry et al., Astrophys. J. 594, 1 (2003)


  \bibitem{s1}T. Clifton, P. G. Ferreira, A. Padilla and C. Skordis,  Phys. Rept. 513, 1  (2012)  
  \bibitem{t1abcd}  V. Acquaviva, C. Baccigalupi, F. Perrotta,  Phys. Rev. D70, 023515 (2004) 
  \bibitem{s2}  R. Klein, M. Ozkan and  D.  Roest,  Phys. Rev. D 93, 044053 (2016) 
    \bibitem{t2abcd}  E. Babichev and  C. Charmousis, JHEP 1408,  106 (2014) 
\bibitem{Jamil:2013yc} 
  M.~Jamil, D.~Momeni and R.~Myrzakulov,
  Eur.\ Phys.\ J.\ C {\bf 73}, no. 3, 2347 (2013)
  doi:10.1140/epjc/s10052-013-2347-4
%\cite{Momeni:2014uwa}
\bibitem{Momeni:2014uwa} 
  D.~Momeni, M.~J.~S.~Houndjo, E.~Güdekli, M.~E.~Rodrigues, F.~G.~Alvarenga and R.~Myrzakulov,
  Int.\ J.\ Theor.\ Phys.\  {\bf 55}, no. 2, 1211 (2016)
  doi:10.1007/s10773-015-2762-4
  [arXiv:1412.4672 [gr-qc]].
\bibitem{s12}  A. Nicolis, R. Rattazzi and E. Trincherini, Phys. Rev. D 79 064036 (2009)
  \bibitem{h1}G. W. Horndeski,  Int. J.
Theor. Phys. 10, 363 (1974) 
\bibitem{h2}C. Deffayet, X. Gao, D. A. Steer and G. Zahariade, 
Phys. Rev. D 84, 064039 (2011) 
  
 \bibitem{h4}G.   Horndeski and J. Wainwright, Phys. Rev. D 16, 1691 (1977) 
 \bibitem{h5}G. Horndeski,  Phys. Rev D 17, 391 (1978)
 
   \bibitem{16}K. Land and J. Magueijo, Phys. Rev. Lett. 95, 071301 (2005)  
  \bibitem{16a}K. Land and J. Magueijo, MNRAS 378, 153  (2007)  
   \bibitem{17} H. K. Eriksen, F. K. Hansen, A. J. Banday, K. M. Gorski and P. B. Lilje, Astrophys. J. 605, 14  (2004) 
     \bibitem{18} L. H. Ford, Phys. Rev. D 40, 967 (1989)  
  \bibitem{19} A. Golovnev, V.  Mukhanov and V.  Vanchurin, JCAP 0806, 009 (2008)  
  \bibitem{21}C. G. Boehmer and T. Harko, Eur. Phys. J. C 50, 423 (2007)  
  \bibitem{21a}T. Koivisto and D. F. Mota, JCAP 0808, 021
(2008)  
  \bibitem{22} J. Beltran Jimenez and A. L. Maroto, Phys. Rev. D 78, 063005 (2008) 
  \bibitem{22a}  J. Beltran Jimenez,
R. Lazkoz and A. L. Maroto, Phys. Rev. D 80, 023004 (2009) 
 \bibitem{12} M. -a. Watanabe, S. Kanno and J. Soda, Phys. Rev. Lett. 102, 191302 (2009) 
 \bibitem{12a}J. Soda, Class.
Quant. Grav. 29, 083001 (2012) 
 \bibitem{12b}A. Maleknejad, M. M. Sheikh-Jabbari and J. Soda, Phys. Rept.
528, 161 (2013) 
 \bibitem{12c} M. Thorsrud, D. F. Mota and S. Hervik, JHEP 1210, 066 (2012) 
 
 
\bibitem{rubin1} V. C. Rubin, E. M. Burbidge, G. R. Burbidge and  K. H. Prendergast , APJ { 141}, 885 (1965)

\bibitem{rubin2}V. C. Rubin and W. K. Ford, Astrophys J. { 159}, 379 (1970)

\bibitem{Zwicky} F. Zwicky, Helv. Phys.  Acta { 6}, 110 (1933)
\bibitem{mond} M. Milgrom, Astrophys. J. 270, 365 (1983)
 \bibitem{mond1}  M. Milgrom, Astrophys. J. 270, 371 (1983)
\bibitem{Moffat1} J. W. Moffat, JCAP 0603 004 (2006).

\bibitem{MoffatRahvar1} J. W. Moffat and S. Rahvar, MNRAS, {  436}, 1439 (2013) 

\bibitem{MoffatRahvar2} J. W. Moffat and S. Rahvar, MNRAS, {  441}, 3724 (2014) 

\bibitem{BrownsteinMoffat1} J. R. Brownstein and J. W. Moffat, MNRAS {  367}, 527 (2006) 
\bibitem{MoffatToth} J. W. Moffat and V. T. Toth, Phys. Rev. {  D91}, 043004 (2015) 



\bibitem{BrownsteinMoffat2} J. R. Brownstein and J. W. Moffat, MNRAS, {  382}, 29 (2007) 
 
\bibitem{Moffat5} J. W. Moffat, Eur. Phys. J. C { 75}, 175 (2015) 

\bibitem{Moffat6} J. W. Moffat,  Eur. Phys. J. C { 75}, 130 (2015) 
 



 
   \bibitem{mirf} J.  R. Mureika, J. W. Moffat and M.  Faizal,  Phys. Lett. B 757, 528 (2016) 

   
   
\bibitem{t1} R. C. Tolman,   Phys.  Rev. 55,  364  (1939)
\bibitem{t2}  J. R. Oppenheimer  and G. M. Volkoff,  Phys. Rev. 55, 374 (1939)
   
   \bibitem{g0}K.~Nordtvedt,  Phys.  Rev.  169, 1014 (1968)
   \bibitem{g1}  K.~Nordtvedt,  Phys. Rev. 169, 1017 (1968)
   \bibitem{g2} K.~Nordtvedt,  Phys. Rev. 170, 1186 (1968)
   
\bibitem{poly12}A. M. Oliveira, H. E. S. Velten, J. C. Fabris and  L. Casarini
Phys. Rev. D 92, 044020 (2015)   
   
   
\bibitem{eos1}
D. Page and S. Reddy,  Annu. Rev. Nucl. Part. Sci. 56, 327 (2006)

\bibitem{eos2}
J. M. Lattimer,  Annu. Rev. Nucl. Part. Sci. 62, 1 (2012)
\bibitem{eos3}
J. M. Lattimer,   AIP Conf. Proc. 1645, 61 (2015)


\bibitem{poly}
M. Prakash, Equation of state, in The Nuclear Equation
of State, ed. by A. Ausari, L. Satpathy (World Scientific,
Singapore, 1996).
\bibitem{method}
W. H. Press, S. A. Teukolsky, W. T. Vetterling and  B. P.
Flannery, Numerical Recipes, The Art of Scientific Computing,  (Cambridge University Press, 2007) 
\bibitem{Oliveira:2015lka} 
  A.~M.~Oliveira, H.~E.~S.~Velten, J.~C.~Fabris and L.~Casarini,
  %``Neutron Stars in Rastall Gravity,''
  Phys.\ Rev.\ D {\bf 92}, no. 4, 044020 (2015)
  doi:10.1103/PhysRevD.92.044020
  [arXiv:1506.00567 [gr-qc]].
\bibitem{Buchdahl}
 H.A. Buchdahl,  Phys. Rev. 116, 1027 (1959)
\bibitem{bound}
A.~Fuzfa, J.~M.~Gerard and D.~Lambert, Gen. Rel. Grav. 34, 1411 (2002)


   \bibitem{co}P.  S. Joshi and  D.   Malafarina,   Eur. Phy.  J. C  75, 596 (2015)
\bibitem{Schlogel:2016xtf} 
  S.~Schlögel,  arXiv:1610.03622 [gr-qc].
\bibitem{2k}K. Yagi and N.  Yunes, 	Phys. Rev. D 91, 103003 (2015)
\bibitem{1k}R. Sharma andS. D. Maharaj,  J. Astrophys.
Astron. 28, 133 (2007) 
 \bibitem{lense1}
 A. F. Boden,  M. Shao and  D.  Van Buren,  Astrophys.  J. 
 502, 538 (1997) 
 \bibitem{lense2} J.Miralda-Escude,  Astrophys.  
 J. Lett. 470, L113 (1996) 
\bibitem{xray} T. S. Olson, Phys. rev. C 63, 015802 (2001) 
\bibitem{quescent} E. F. Brown, L. Bildsten and R. E. Rutledge, Astrophys. J. 504,  L95 (1998)
\bibitem{reis}  F. Ficek,  Class. Quant. Grav. 32,  235008 (2015)
\bibitem{reis1}   A. K. Ahmed, U. Camci and M. Jamil,  Class. Quant. Grav. 33,  215012 (2016)

\end{thebibliography}
\end{document}